\documentclass[english,preprintnumbers,amsmath,amssymb,aps,review,nofootinbib]{revtex4}
\usepackage[T1]{fontenc}
\usepackage[latin9]{inputenc}
\usepackage{color}
\usepackage[linktocpage=true,plainpages=false,pdfpagelabels=false]{hyperref}
\usepackage{doi}
\usepackage{textcomp}
\usepackage{amsmath}
\usepackage{amssymb}
\usepackage{esint}
\definecolor{linkcolor}{rgb}{0.6,0,0}
\definecolor{citecolor}{rgb}{0,0.5,0}
\definecolor{urlcolor}{rgb}{0,0,1}
\hypersetup{colorlinks, linkcolor={linkcolor},
    citecolor={citecolor}, urlcolor={urlcolor}}

\makeatletter

\@ifundefined{textcolor}{}
{%
 \definecolor{BLACK}{gray}{0}
 \definecolor{WHITE}{gray}{1}
 \definecolor{RED}{rgb}{1,0,0}
 \definecolor{GREEN}{rgb}{0,1,0}
 \definecolor{BLUE}{rgb}{0,0,1}
 \definecolor{CYAN}{cmyk}{1,0,0,0}
 \definecolor{MAGENTA}{cmyk}{0,1,0,0}
 \definecolor{YELLOW}{cmyk}{0,0,1,0}
}

\usepackage{dcolumn}
\usepackage{bm}
\usepackage{amsfonts}\usepackage{babel}
\providecommand{\U}[1]{\protect\rule{.1in}{.1in}}

\@ifundefined{textcolor}{}{
\definecolor{BLACK}{gray}{0}
 \definecolor{WHITE}{gray}{1}
 \definecolor{RED}{rgb}{1,0,0}
 \definecolor{GREEN}{rgb}{0,1,0}
 \definecolor{BLUE}{rgb}{0,0,1}
 \definecolor{CYAN}{cmyk}{1,0,0,0}
 \definecolor{MAGENTA}{cmyk}{0,1,0,0}
 \definecolor{YELLOW}{cmyk}{0,0,1,0}
 }

\usepackage{babel}
\usepackage{ragged2e}

\makeatother

\usepackage{babel}
\begin{document}

\title{General Relativity à la string: a progress report\footnote{Research sponsored by the National Science Foundation under Grants
No. GP-30799X to Princeton University and GP-40768X to the Institute
for Advanced Study.}\footnote{This paper was originally published as: T.~Regge and C.~Teitelboim, `General
Relativity à la string: a progress report,' in \emph{Proceedings
of the First Marcel Grossmann Meeting (Trieste, Italy, 1975)}, ed.
by R.~Ruffini, 77\textendash 88, North-Holland, Amsterdam, 1977. Several colleagues have made the point that this reference is hard to access, and have suggested that it should be reprinted in the arXiv to make it available. This is the purpose of the present text. Especial thanks are expressed to Sergey Paston and Anton Sheykin for taking the initiative and going the effort of transforming the ``camera ready'' text of more than forty years ago to TeX. The kind help of Alfredo P\'{e}rez on this front is also gratefully acknowledged. Recognition is expressed to Georgi Dvali for giving the decisive push when referring to the article as ``the paper that does not exist.''}  }

\author{Tullio Regge}

\affiliation{ Institute for Advanced Study, Princeton, New Jersey 08540}

\author{Claudio Teitelboim}


\affiliation{Joseph Henry Laboratories, Princeton University, Princeton, New Jersey 08540}


\begin{abstract}
Preliminary results on a canonical formulation of general relativity
based on an analogy with the string model of elementary particles
are presented. Rather than the metric components, the basic fields
of the formalism are taken to be the functions describing the embedding
of four dimensional spacetime in a ten or possibly higher dimensional
manifold. So far, the main drawback of the formalism is that the generator
of normal deformations (``fourth constraint'') cannot be written
down in closed form. The present approach is compared and contrasted
with the usual one and with the canonical description of the relativistic
string.
\end{abstract}
\maketitle

It is our intention in this note to analyze some formal analogies
existing in different relativistic systems and to assess some of the
current difficulties in the canonical formalism for general relativity.
In particular we shall compare two different approaches to general
relativity (the conventional one (\ref{A}) \cite{1,2} and a new
one based on the notion of external variables (\ref{B})) with the
string model of elementary particles \cite{3}. 

\section{General Relativity: Conventional Formalism}

\label{A} In the usual canonical formalism for Einstein's theory
of gravitation developed by Dirac \cite{1} and Arnowitt, Deser and
Misner (ADM) \cite{2} the starting point is the Hilbert action 
\begin{align}
S= & \int\left(-^{\left(4\right)}g\right)^{1/2}\,{}^{\left(4\right)}R\,d\,^{4}x\label{1}
\end{align}
which is regarded as a functional of the metric tensor $g_{\mu\nu}(x)$,
$x\in R^{4}$.

It is an important feature of the Hilbert action that by adding a
suitable divergence to the integrand in \eqref{1} one can switch
to an alternate action density \textemdash{} the Dirac \textemdash{}
ADM \textemdash lagrangian density

\begin{align}
\mathcal{L}= & \left(-^{\left(4\right)}g\right)^{1/2}\,{}^{\left(4\right)}R+\frac{\partial \cal{V}^{\alpha}}{\partial x^{\alpha}}\label{2}
\end{align}
which contains no second time derivatives of the $g_{\mu\nu}$ and
furthermore contains no first time derivatives of the $g_{0\mu}$.

The $g_{0\mu}$ have then vanishing conjugate momenta and enter the
theory as arbitrary functions. At this stage the remaining degrees
of freedom are thus those represented by the spatial metric components
$g_{ij}$ and their conjugates $\pi^{ij}$ . The fields $g_{ij}$,
$\pi^{ij}$ are however not independent, but they are restricted by
the constraint equations\footnote{The ``weak equality'' symbol is used to emphasize that $\mathcal{H}_{\bot}$
and $\mathcal{H}_{i}$ have non-vanishing Poisson brackets with the
canonical variables of the theory. The vertical slash denotes covariant
differentiation in the spatial metric $g_{ij}$. Spacetime covariant
derivatives are indicated by a semicolon. The letter $R$ denotes
the spatial curvature and $g$ is the determinant of the spatial metric.
To avoid confusion some spacetime quantities carry an upper left index
$(4)$ as in \eqref{1}.} \begin{subequations}\label{3} 
\begin{align}
 & \mathcal{H}_{\bot}=g^{-1/2}\left(\pi_{ij}\pi^{ij}-\frac{1}{2}\left(\pi_{i}^{i}\right)^{2}\right)-g^{1/2}R\approx0,\label{3a}\\
 & \mathcal{H}_{i}=-2\pi_{i}\,^{j}\,_{|j}\approx0.\label{3b}
\end{align}
\end{subequations} Geometrically speaking the $\mathcal{H}_{i}$
generate arbitrary reparametrizations of the spacelike hypersurface
on which the state is defined whereas $\mathcal{H}_{\bot}$ generates
deformations which change the location of the hypersurface in the
ambient spacetime. The fact that the hypersurfaces are embedded in
a common spacetime is expressed through the closure relations \cite{4,5}
\begin{subequations}\label{4} 
\begin{align}
 & [\mathcal{H}_{\bot}(x),\mathcal{H}_{\bot}(x')]=\left(g^{rs}(x)\mathcal{H}_{s}(x)+g^{rs}(x')\mathcal{H}_{s}(x')\right)\delta_{,r}(x,x'),\label{4a}\\
 & [\mathcal{H}_{r}(x),\mathcal{H}_{\bot}(x')]=\mathcal{H}_{\bot}(x)\delta_{,r}(x,x'),\label{4b}\\
 & [\mathcal{H}_{r}(x),\mathcal{H}_{s}(x')]=\mathcal{H}_{r}(x')\delta_{,s}(x,x')+\mathcal{H}_{s}(x)\delta_{,r}(x,x').\label{4c}
\end{align}
\end{subequations} In this theory it is in principle possible to
fix the gauges by imposing particular coordinate conditions on the
surface and also by fixing the time slicing. The fixation of the spacetime
coordinates amounts therefore to bring in four extra constraints besides
\eqref{3}. After this is done one is left with only two independent
pairs of canonical variables per space point. These degrees of freedom
appear in the weak field approximation as the two polarization states
per wave vector $\tilde{k}$ of a massless spin two graviton propagating
on a flat background.

The practical implementation of the coordinate fixing is unfortunately
frought with difficulties which have prevented so far the construction
of an actual canonical quantum theory of gravity. In the first place
it is not a simple matter to fix the gauge freedom in such a manner
as to ensure a proper parametrization of spacetime through coordinates,
although some of the proposed choices look reasonable \cite{2,6,7}.
A second difficulty is that the reduced Hamiltonian associated to
the coordinate conditions proposed so far cannot be written down in
closed form and usually appears as a highly non-local expression in
the canonical fields. This brings virtually to a halt the construction
of the quantum theory because of the formidable problems of ordering
which must be solved \underline{ex-novo} at each order of perturbation
theory in the expression for the Hamiltonian.

Yet another difficulty arises in the so-called maximal slicing $(\pi^{i}\,_{i}=0)$,
which appears to be the gauge condition most exhaustively investigated
from the point of view of ensuring a proper parametrization of spacetime
\cite{7}. The difficulty in question is that ordering problems appear
here already at the level of interpreting the Poisson brackets of
the basic fields as commutators, because $q$-numbers appear nontrivially
on the right hand side of the commutation relations. Such difficulties
do not arise however for the ADM variables \cite{2,8}, but unfortunately
there is not much evidence that the ADM gauge defines a good system
of spacetime coordinates.

The difficulties mentioned above are by no means exclusive to the
gravitational field and they also appear, for example, in the string
model which bears in many respects a striking analogy with Einstein's
theory of gravitation. In the case of the string, because of the simple
geometrical nature of the model, it is possible to circumvent the
ordering problem by means of the DDF variables \cite{9} as suggested
by the interpretation of the theory in the framework of the dual models
of hadrons.

In what follows we would like to examine the possibility of extending
some of the useful concepts of the string model into general relativity.
Although we have not been successful in this attempt we feel that
the comparative discussion of the two systems is interesting by itself
and leads to useful critical remarks.

\section{The String Model}

\label{B} Here we consider $n+1$ fields $y^{A}(x,t)$, $x\in R$.
The functions $y^{A}$ parametrize a two dimensional surface $V_{2}$
embedded in an $N+1$ dimensional Minkowski space of metric 
\begin{align}
ds^{2}=d\tilde{y}\cdot d\tilde{y}=\eta_{AB}dy^{A}dy^{B}=-(dy^{0})^{2}+\sum\limits _{1}^{N}(dy^{A})^{2}.\label{5}
\end{align}
The two dimensional surface is spanned by the motion of the (one dimensional)
string in the $N+1$ dimensional space.

The action for the system is taken to be 
\begin{align}
S= & \int\,\left(-^{\left(2\right)}g\right)^{1/2}\,dx\,dt\label{6}
\end{align}
where $\left(-^{\left(2\right)}g\right)^{1/2}\,dx\,dt$ is the area
element on $V_{2}$. The string is assumed to have a finite length
and one has to impose Poincar$\acute{\text{e}}$ invariant boundary
conditions at its ends in order to obtain a relativistic theory. The
boundary conditions imply that the endpoints move transversally with
the speed of light. The canonical formalism based on \eqref{6} leads
to a vanishing canonical Hamiltonian (due to the time reparametrization
invariance of \eqref{6}) and to constraints of the form \begin{subequations}\label{7}
\begin{align}
 & \mathcal{H}_{1}=\tilde{\pi}\cdot\frac{\partial\tilde{y}}{\partial x}\approx0,\label{7a}\\
 & \mathcal{H}_{\bot}=\frac{1}{2}\left|\frac{\partial\tilde{y}}{\partial x}\right|^{-1}\left(\tilde{\pi}^{2}+\left(\frac{\partial\tilde{y}}{\partial x}\right)^{2}\right)\approx0.\label{7b}
\end{align}
\end{subequations} The functions \eqref{7} admit again the geometrical
interpretation of generating tangential and normal deformations of
the string. They satisfy closure relations analogous to \eqref{4},
namely \cite{8} \begin{subequations}\label{8} 

\begin{align}
[\mathcal{H}_{\bot}(x),\mathcal{H}_{\bot}(x')]=&\left(\left|\frac{\partial\tilde{y}}{\partial x}\right|^{-2}\!\!\!\!(x)\,\mathcal{H}_{1}(x)+\left|\frac{\partial\tilde{y}}{\partial x}\right|^{-2}\!\!\!\!(x')\,\mathcal{H}_{1}(x')\right)\delta'(x,x')+\nonumber\\
&+2\left(\left|\frac{\partial\tilde{y}}{\partial x}\right|^{-3}\!\!\!\!(x')\,\mathcal{H}_{\bot}(x)\mathcal{H}_{1}(x)+\left|\frac{\partial\tilde{y}}{\partial x}\right|^{-3}\!\!\!\!(x')\,\mathcal{H}_{1}(x')\right)\delta'(x,x'),\label{8a}\\
  [\mathcal{H}_{1}(x),\mathcal{H}_{\bot}(x')]=&\mathcal{H}_{\bot}(x)\delta'(x,x'),\label{8b}\\
  [\mathcal{H}_{1}(x),\mathcal{H}_{1}(x')]=&(\mathcal{H}_{1}(x)+\mathcal{H}_{1}(x'))\delta'(x,x').\label{8c}
\end{align}
\end{subequations}

We note that the only difference between \eqref{8} and \eqref{4}
is the presence of the term quadratic in the constraints on the right
hand side of \eqref{8a}. This term has however weakly vanishing brackets
with everything, which means that \eqref{7} still ensures that all
the strings are embedded in a common two dimensional Riemannian surface.

In GGRT \cite{3} the problem of accounting for the constraints and
fixing the coordinate system on the surface spanned by the string
is solved by introducing a system of null surfaces $y^{0}-y^{1}=t$
in $R^{N+1}$ which reduces the problem to dealing with $N-1$ independent
modes per point on the string\footnote{Null surfaces have been introduced to analyze the dynamics of gravity
by Aragone and Gambini \cite{10} and Kaku \cite{11}. There is however
no analog in the discussion given by those authors of an ambient flat
space which is heavily relied upon in the string model.}. It is also possible to introduce a more conventional spacelike gauge
\cite{12} $y^{0}=t$. In the latter case the Dirac brackets of the
basic fields are given typically by expressions of the form \begin{subequations}\label{9}
\begin{align}
[\alpha_{m}^{A},\alpha_{n}^{B}]=m\delta_{m,-m}\delta^{AB}+\sum\limits _{M\neq0}\frac{mn}{M}\frac{1}{(p^{0})^{2}}\alpha_{m-M}^{A}\alpha_{n+M}^{B}\label{9a}
\end{align}
where 
\begin{align}
y^{A}(x,t) & =q^{A}+p^{A}t+i\sum\limits _{n\neq0}\frac{1}{n}\alpha_{n}^{A}\cos(nx)e^{-int}\label{9b}
\end{align}
\end{subequations}

Equation \eqref{9} shows that the fields $y^{A}$, $\pi_{A}$ are
related to the fundamental canonical variables of the theory by a
non-elementary expression. It is in fact extremely hard to approach
the quantization procedure by considering the $y^{A}$ as operators
and \eqref{9a} as a commutation relation because of the ordering
problem. A better approach is to consider the DDF operators which
appear in the integral form 
\begin{align}
D_{n}^{A}=\frac{1}{2}\int\limits _{\pi}^{2\pi}\frac{dy^{A}(0,t)}{dt}\text{exp}\left(n(\tilde{k}\cdot\tilde{p})^{-1}\tilde{k}\cdot\tilde{y}(0,t)\right)dt
\end{align}
(here $\tilde{k}$ is an arbitrary null vector) and which obey a simple
algebra. The whole string model can be built upon a systematic exploitation
of this algebra.

The underlying pseudo Euclidean structure of $R^{N+1}$ is necessary
for the use of the DDF operators in that form it follows that there
are orthonormal coordinates $x,t$ such that the equations of motion
can be explicitly solved in the form 
\begin{align}
y^{A}(x,t)=f^{A}(t-x)+f^{A}(t+x),\label{11}
\end{align}
an equation which is crucial in defining the Fourier transform used
by DDF.

A solution similar to \eqref{11} is of course not available in general
relativity but it is nevertheless of interest to investigate what
happens if one tries to cast general relativity in a string-like form,
which we pass to do now.

\section{General Relativity À la String}

By analogy with the string model we postulate here that ordinary curved
spacetime $V_{4}$ is embedded in some Minkowski space $R^{N+1}$
with a sufficiently high dimensionality $N\geq9$ so as to be able
to accommodate locally a generic four-dimensional pseudo Riemannian
manifold. We thus consider the spacetime $V_{4}$ as a ``trajectory''
swept by a three dimensional string in $R^{N+1}$.

The key difference between the present formalism and the usual approach
described in Section \ref{A} above, is that the metric components
$g_{\mu\nu}(x)$ are no longer the basic variables but, rather, they
are regarded now as derived objects constructed from the functions
$y^{A}(x^{0},x^{1},x^{2},x^{3})$ determining the (time dependent)
embedding of $V_{3}$ in $R^{N+1}$.

The metric tensor is thus given by 
\begin{align}
g_{\mu\nu}(x)=\tilde{y}_{,\mu}\cdot\tilde{y}_{,\nu}=\eta_{AB}\frac{\partial\tilde{y}^{A}}{\partial x^{\mu}}\frac{\partial\tilde{y}^{B}}{\partial x^{\nu}}\label{12}
\end{align}
with $\eta_{AB}=\text{diag}(-1,1,\ldots,1)$, $A,B,\ldots=0\ldots N$. 

We shall use the same action as in \ref{A}, namely 
\begin{align}
S[y]=\int\mathcal{L}\,d^{4}\,x\label{13}
\end{align}
where $\mathcal{L}$ is the Dirac-{}-ADM lagrangian density appearing
in \eqref{2}, regarded this time as a functional of the $y^{A}$
through \eqref{12}. The fact that $\mathcal{L}$ contains no time
derivatives of the $g_{0\mu}$ implies that only first time derivatives
of the $y^{A}$ enter into the action \eqref{13}. Eq. \eqref{12}
shows that $y_{,0}^{A}$ can enter $\mathcal{L}$ through $g_{0\alpha}$
only). As solely first time derivatives of the $y^{A}$ appear in
the action we see that we are still dealing with a system that can
be put in canonical form by standard methods. We have already paid
however, a stiff price by introducing the external variables $y^{A}$
, namely, we have to retain all the fields instead of being able to
eliminate four of them (the $g_{0}$) at an early stage as was done
in \ref{A}.

A worse feature is that requiring the action \eqref{13} to be stationary
under arbitrary variations of the $y^{A}$ does not reproduce the
equations of motion of general relativity

\begin{align}
(\text{Einstein tensor})^{\alpha\beta} & =G^{\alpha\beta}=0,\label{14}
\end{align}
but gives rather the weaker set 
\begin{align}
G^{\alpha\beta}\tilde{y}_{;\alpha\beta}=0.\label{15}
\end{align}
Equations \eqref{15} are the analog of the string equations 
\begin{align}
g^{\alpha\beta}\tilde{y}_{;\alpha\beta}=0. & \quad(\text{String})\label{16}
\end{align}
in which case $\alpha$ and $\beta$ refer to the two dimensional
spanned by the string.

Equations \eqref{14} do not imply $G^{\alpha\beta}=0$ due to the
identities 
\begin{align}
\tilde{y}_{;\alpha\beta}\cdot\tilde{y}_{,\gamma}=0.\label{17}
\end{align}
which show that in the generic case only six among the $N+1$ equations
are independent. We note in passing that the identities \eqref{17}
avoid the paradoxical implication $g^{\alpha\beta}=0$ in \eqref{16}.
The difficulty of having only six independent equations in \eqref{15}
instead of the full Einstein set is not unsurmountable and could be
circumvented by imposing in an ad-hoc fashion the additional constraints
\begin{align}
G_{\bot\alpha}=0\label{18}
\end{align}
where the symbol $\bot$ refers to the unit normal to $V_{3}$ lying
in $V_{4}$ and $a=\bot,1,2,3$. Examination of the canonical formalism
for the external variables shows in fact that one may expect \eqref{18}
not to be an entirely unreasonable addition to the equations \eqref{15}. 

\section{Canonical Formalism for External Variables}

We start from the Dirac-ADM Lagrangian density \eqref{13}, which
written down in detail reads 
\begin{align}
\mathcal{L}= & g^{1/2}N(R+K_{ab}K^{ab}-(K_{a}^{a})^{2}),\label{19}
\end{align}
where $R$ is the curvature scalar of $V_{3}$ and where the extrinsic
curvature $K_{ab}$ of $V_{3}$ with respect to $V_{4}$ is given
by 
\begin{align}
K_{ab}=(2N)^{-1}(-\dot{g}_{ab}+N_{a|b}+N_{b|a}).\label{20}
\end{align}
The symbols $N$ and $N_{a}$ stand for the lapse and shift functions
\begin{align}
N=\left(-^{\left(4\right)}g^{00}\right)^{1/2},\ N_{a}=g_{0a}.\label{21}
\end{align}
A dot denotes differentiation with respect to $x^{0}$. The Lagrangian
density \eqref{19} is expressed as a functional of the $y^{A}$ by
means of \eqref{12}, \eqref{20} and \eqref{21}.

The canonical momenta are defined by 
\begin{align}
\tilde{\pi}(x)=\frac{\delta}{\delta\dot{\tilde{y}}(x)}\int d^{3}x'\mathcal{L}(x')\label{22}
\end{align}
which gives, after some calculation\footnote{Note added (2016): Actually the right-hand side of (\ref{23}) is nothing but the Lagrangian density obtained by dropping the factor $N$ in (\ref{19}). See \cite{BP1}.}, 
\begin{align}
\tilde{\pi}= & g^{1/2}\left(-2G_{\bot\bot}\tilde{n}+2(K^{ab}-K_{m}^{m}g^{ab})\tilde{y}_{|ab}\right)\label{23}
\end{align}
Here $\tilde{n}$ denotes the unit normal to $V_{3}$ lying in $V_{4}$:
\begin{align}
\tilde{n}=N^{-1}\left[\dot{\tilde{y}}-(\dot{\tilde{y}}\cdot{\tilde{y}}^{|i}{\tilde{y}}_{,i})\right]\label{24}
\end{align}
and $G_{\bot\bot}$ is the double projection of the Einstein tensor
along $\tilde{n}$ : 
\begin{align}
-2G_{\bot\bot}=K_{ab}K^{ab}-(K_{m})^{2}-R.\label{25}
\end{align}
The normal \eqref{24} satisfies the normalization condition, 
\begin{align}
\tilde{n}\cdot\tilde{n}=-1,\label{26}
\end{align}
and the extrinsic curvature is related to $\tilde{n}$ by 
\begin{align}
K_{ab}=\tilde{n}\cdot\tilde{y}_{|ab}.\label{27}
\end{align}
Now we note that the six vectors $\tilde{y}_{|ab}$ are perpendicular
to $V_{3}$ (this is just the $V_{3}$ version of the identities \eqref{17}).
Also the normal $\tilde{n}$ is perpendicular to $V_{3}$. It thus
follows that the three components of $\tilde{\pi}$ on $V_{3}$ vanish.
We then get the three constraints 
\begin{align}
\mathcal{H}_{i}=\tilde{\pi}\cdot\tilde{y}_{,i}\approx0,\label{28}
\end{align}
which are the analog of \eqref{7a} for the string. The $\mathcal{H}_{i}$
defined by \eqref{28} generate reparametrizations on $V_{3}$ and
they satisfy consequently the closure relations \eqref{4c}. It follows
from \eqref{28}, for example that $y^{A}$ and $\pi_{A}$ transform
as scalars and scalar densities respectively under changes of coordinates
in $V_{3}$, which was of course to be expected.

The fourth constraint (analog to \eqref{7b}) for the string) is obtained
in principle by solving \eqref{23} as a system of nonlinear algebraic
equations for ${n}^{A}$ as a function of ${\pi}_{A}$ and $y^{A}$
and imposing afterwards the normalization condition \eqref{26}. In
the case of the string this procedure yields \eqref{7b}. In fact
the string analog of \eqref{23} reads simply 
\begin{align}
\tilde{\pi}=\left|\frac{\partial\tilde{y}}{\partial x}\right|\tilde{n}\quad(\text{string}),\label{29}
\end{align}
which upon squaring and using \eqref{26}, gives \eqref{7b}. The
solution of \eqref{23} is however considerably harder and there seems
to be no way of obtaining a simple closed form for 
\begin{align}
\tilde{n}=\tilde{n}(\tilde{y},\tilde{\pi}).\label{30}
\end{align}
This problem might be circumvented to some extent with the help of
the additional constraints \eqref{18} which we pass to discuss now.

As we mentioned before, even if we imagine having the solution (30),
the formalism does not reproduce Einstein's theory. In fact, if we
count the number of independent degrees of freedom of the theory we
find: $2(N+1)-4$ (first class constraints) $-4$ (gauge conditions)
$=2(N-3)$. That is we have $N-3$ degrees of freedom per point, which
recalling that $N\geq9$ is at least an excess of four over the required
number of two for general relativity.

We see therefore that even if we bring $N$ down to its minimum value
of 9, as we shall do tentatively from now on, we need four additional
first class constraints besides the $\mathcal{H}_{\mu}$. It is quite
reasonable to take these new constraints to be \eqref{18}. In fact
the $G_{\bot\mu}$ are constructed from the $g_{ab}$ and $K_{ab}$
only, which means that they can in principle be expressed, via \eqref{30},
as functions of the canonical variables $\tilde{y}$, $\tilde{\pi}$.

Now, if we are going to impose the constraints \eqref{18}, we need
to solve \eqref{23} for $\tilde{n}$ only when $G_{\bot\bot}=0$.
This will result in changing the constraints by linear combinations
of themselves and will therefore not change the dynamics of the system.

When $G_{\bot\bot}=0$ \eqref{23} can be written as 
\begin{align}
{\pi}^{A}\approx W_{B}^{A}n^{B},\label{31}
\end{align}
with 
\begin{align}
W_{B}^{A}= & 2g^{1/2}(g^{ad}g^{bc}-g^{ab}g^{cd})y_{|ab}^{A}y_{B|cd}.\label{32}
\end{align}
The matrix $W$ defined by \eqref{32} regarded as a mapping of $R^{10}$
onto $R^{10}$ does not have an inverse because it maps to zero the
three vectors $\tilde{y}_{,i}(i=1,2,3)$. However when restricted
to the sub-space orthogonal to the $\tilde{y}_{,i}$, $W$ will have
an inverse in the generic case. Let us denote that inverse by $M$.
The matrix $M$ is therefore defined as giving that solution of \eqref{31},
\begin{align}
n^{B}=M_{A}^{B}\pi^{A}\label{33}
\end{align}
which satisfies 
\begin{align}
\tilde{n}\cdot\tilde{y}_{,i}=0.\label{34}
\end{align}
It follows from \eqref{32} that $M$ is constructed from the ${y}^{A}$
and their derivatives and that $M_{AB}=\eta_{AC}M_{B}^{C}$ is symmetric.
The eight constraints of the theory can then be expressed in terms
of $M$ as follows \begin{subequations}\label{35} 
\begin{align}
-2G_{\bot\bot} & =K_{ab}K^{ab}-(K_{m}^{m})^{2}-R\approx\frac{1}{2}g^{-1/2}M_{AB}\pi^{A}\pi^{B}-R\approx0,\label{35a}\\
-G_{\bot i} & =(K_{i}^{k}K_{m}^{m}\delta_{i}^{k})_{|k}\approx(M_{AB}\pi^{B})_{,i}y^{A|m}\,_{|m}-(M_{AB}\pi^{B})_{,m}y^{A|m}\,_{|i}\approx0,\label{35b}\\
\mathcal{H}_{\bot} & =g^{1/2}(\tilde{n}^{2}+1)\approx g^{1/2}\left((M^{2})_{AB}\pi^{A}\pi^{B}+1\right)\approx0,\label{35c}\\
\mathcal{H}_{i} & =\tilde{\pi}\cdot\tilde{y}_{,i}\approx0.\label{35d}
\end{align}
\end{subequations} The essential problem at this point is to prove
that the eight constraints \eqref{35} are first class. It does not
seem possible to do this without knowing more about the form of $M_{AB}$.
We plan to investigate this matter in the future.

If the constraints \eqref{35} are indeed first class their compatibility
is ensured and the theory is consistent. Furthermore we are then sure
that we are dealing exactly with Einstein's equations because the
only way in which $G_{\bot\mu}$ can vanish on every three dimensional
space like hypersurface of $V_{4}$ is that all ten equations $G_{\alpha\beta}=0$
hold. On the other hand, if the system \eqref{35} is not first class
we would be merely selecting by means of \eqref{35a}, \eqref{35b}
special coordinates on $V_{4}$ (i.e., fixing the gauge) instead of
reducing the number of physical degrees of freedom of the theory.
The formalism would not reproduce Einstein's theory in that case. 

\section*{Final Remarks}

The theory as we have presented it here is not complete but we feel
it deserves further investigation. It is quite possible that the actual
value of $N$ is not relevant in a final, as yet hypothetical, complete
form. We must keep in mind in this connection that the possibility
of embedding a four dimensional manifold in $R^{10}$ holds only in
a very local sense and that non-trivial problems are already encountered
in trying to embed globally a smooth two dimensional manifold \cite{13}
in $R^{3}$. However the existence of the variables $y^{A}$ gives
us more freedom to construct field variables which do not exist in
the conventional theory and which could possibly lead to a canonical
formulation of general relativity different from the conventional
one. In this sense it could be interesting to try to find the analog
of the DDF operators for the string model. Unfortunately we have not
been able as yet to obtain any definite result along this direction.

\begin{acknowledgments}
The authors are indebted to Professors Abdus Salam and Gallieno Denardo
for their kind hospitality at Trieste. One of us (C.T.) would also
like to thank John Wheeler for much encouragement.

\end{acknowledgments}

\end{document}